\documentclass[useAMS,usenatbib,onecolumn]{mn2e}

\usepackage{times}
\usepackage{graphicx}
\usepackage{amsmath}
\usepackage{epsf}
\usepackage{natbib}
\bibpunct{(}{)}{;}{a}{}{,} 

\def\pdeg{\ifmmode $\setbox0=\hbox{$^{\circ}$}\rlap{\hskip.11\wd0 .}$^{\circ}
          \else \setbox0=\hbox{$^{\circ}$}\rlap{\hskip.11\wd0 .}$^{\circ}$\fi}
\def\arcs{\ifmmode {^{\scriptscriptstyle\prime\prime}}
          \else $^{\scriptscriptstyle\prime\prime}$\fi}
\def\arcm{\ifmmode {^{\scriptscriptstyle\prime}}
          \else $^{\scriptscriptstyle\prime}$\fi}
\newdimen\sa  \newdimen\sb
\def\parcs{\sa=.07em \sb=.03em
     \ifmmode \hbox{\rlap{.}}^{\scriptscriptstyle\prime\kern -\sb\prime}\hbox{\kern -\sa}
     \else \rlap{.}$^{\scriptscriptstyle\prime\kern -\sb\prime}$\kern -\sa\fi}
\def\parcm{\sa=.08em \sb=.03em
     \ifmmode \hbox{\rlap{.}\kern\sa}^{\scriptscriptstyle\prime}\hbox{\kern-\sb}
     \else \rlap{.}\kern\sa$^{\scriptscriptstyle\prime}$\kern-\sb\fi}
\font\like=cmsy10 at 10pt
\font\xxxx=cmr10 at 10pt
\def\lea{\mathrel{\raise .4ex\hbox{\rlap{$<$}\lower 1.2ex\hbox{$\sim$}}}}
\def\gea{\mathrel{\raise .4ex\hbox{\rlap{$>$}\lower 1.2ex\hbox{$\sim$}}}}
\let\lsim=\lea

\def\leaderfil{\leaders\hbox to 5pt{\hss.\hss}\hfil}
\def\doubleline{\vskip 3pt\hrule \vskip 1.5pt \hrule \vskip 5pt}

\def\reff@jnl#1{{\rm#1\/}}
\def\apj{\reff@jnl{ApJ}}       
\def\apjs{\reff@jnl{ApJS}}     
\def\apjl{\reff@jnl{ApJL}}     
\def\aaps{\reff@jnl{A\&AS}}    
\def\mnras{\reff@jnl{MNRAS}}   
\def\prd{\reff@jnl{Phys.\ Rev.\ D}}    

\def\simpropto{\,{\lower3.0pt\hbox{$\propto$}\atop\raise1.0pt\hbox{$\sim$}}\,}

\newcommand{\vW}{\mathbf{W}}
\newcommand{\vK}{\mathbf{K}}

\newcommand{\beq}{\begin{equation}}
\newcommand{\eeq}{\end{equation}}

\newcommand{\tC}{\widetilde{C}}
\newcommand{\tN}{\widetilde{N}}

\newcommand{\n}{{\bf{n}}}

\newcommand{\be}{\begin{equation}}
\newcommand{\ee}{\end{equation}}
{\begin{enumerate}\setlength{\itemsep}{0mm}}%
{\end{enumerate}}
{\begin{enumerate}\setlength{\itemsep}{0mm}}%
{\end{enumerate}}
\begin{document}

\title[{\tt XFaster} likelihood]{Performance of {\tt XFaster} likelihood in real CMB experiments}
\author[G. Rocha et al.]
{G. Rocha,$^{1,2}$ C. R. Contaldi,$^{3}$ L. P. L. Colombo,$^{4}$ J. R. Bond,$^{5}$ K. M. G\'orski,$^{1,6}$ and \newauthor C. R. Lawrence$^{1}$ \\
$^{1}$ Jet Propulsion Laboratory, California Institute of Technology, 4800 Oak Grove Drive, Pasadena CA 91109, U.~S.~A. \\
$^{2}$ Department of Physics, California Institute of Technology, Pasadena, 91125, U.~S.~A. \\
$^{3}$ Department of Physics, Imperial College, University of London, South Kensington Campus, London SW7 2AZ, U.~K. \\
$^{4}$ Department of Physics and Astronomy, University of Southern California, University Park Campus, Los Angeles CA 90089, U.~S.~A.  \\
$^{5}$ The Canadian Institute for Theoretical Astrophysics, CITA, University of Toronto, 60 St. George Street, Toronto, Ontario, M5S 3H8, Canada \\
$^{6}$ Warsaw University Observatory, Aleje Ujazdowskie 4, 00478 Warszawa, Poland}

\date{Accepted ---. Received ---.; in original form \today}

\date{\today} 
\maketitle

\label{firstpage}

\maketitle

\begin{abstract}

We assess the strengths and weaknesses of several likelihood formalisms, including the {\tt XFaster} likelihood.  
We compare the performance of the {\tt XFaster} likelihood to that of the Offset Lognormal Bandpower likelihood on simulated data for the Planck satellite. 
 Parameters estimated with these two likelihoods are in good agreement.  
 The advantages of the {\tt XFaster} likelihood can therefore be realized without compromising performance.
   
\end{abstract}    
  


\section{Introduction}
\label{intro}

The temperature and polarization anisotropies of the Cosmic Microwave Background (CMB) contain a wealth of cosmological information.  In extracting this information from measurements of the CMB, the likelihood function $\hbox{\like L}(C_\ell) = P(\hbox{\xxxx data}|C_\ell)$, where $C_\ell$ is the theoretical spectrum for some cosmological model, plays an important role.  For Gaussian fluctuations, $\hbox{\like L}(C_\ell)$ is given by a Multivariate Gaussian of the observed data (section~\ref{multi-gauss}).  For low resolution data ($\ell\lsim10^2$, defined more precisely later) in the usual spherical harmonic expansion, it is computationally feasible to evaluate this directly.  For high resolution data, it is not, and computationally tractable approximations are required.  Fortunately, the large number of independent samples of the universe at high resolution, coupled with a much more nearly diagonal covariance matrix, mean that approximations exist that are accurate as well as fast.   The subject of this paper is the performance of one such high $\ell$ method, the ``{\tt XFaster} likelihood.''   

A successful high $\ell$ method must account correctly for correlations induced in the angular power spectra by partial sky coverage, non-uniform noise, and non-zero beamwidths, as well as the temperature and polarization cross power.  A number of approaches for high-$\ell$ have been proposed.  For temperature alone, these include the Gaussian, Offset Lognormal, Equal Variance \citep {DABJK00}, WMAP \citep {DA-Ver03}, and  SCR \citep {smith06} likelihood approximations.  For temperature and polarization together, these include the Offset Lognormal Bandpower, Hamimeche and Lewis \citep {HL_like}, and {\tt XFaster} \citep {XFaster-like2} likelihoods.

This paper is organized as follows. In Section~\ref{multi-gauss} we describe the the Multivariate Gaussian likelihood both in pixel and harmonic space, give an overview of existing high $\ell$ approximations (Section~\ref{like-approx}) and their limitations. An account of their performances applied to simulated Planck data has been given in \citep{XFaster-like2}.  We further test the performance of {\tt XFaster} likelihood implemented in a new, modified version of the publicly available software \texttt{CosmoMC} code\footnote {http://cosmologist.info/cosmomc/} (\citet{Lewis:2002ah}) by applying it to estimation of cosmological parameters, and compare it to the Offset Lognormal Bandpower likelihood.   We further test our approach to tackle the asymmetry of the beams by comparing parameters estimated for Planck simulated data convolved with a symmetric and a asymmetric beam using both likelihood approaches.

\section{Likelihood for a Gaussian sky}
\label{multi-gauss}

Pixel temperature fluctuations $T({\hat\n})$ (Stokes $I$) on the celestial sphere can be expanded in terms of spherical harmonics $Y_{\ell m}$ as
\begin{equation}
 T({\hat\n}) = \sum_{\ell m} a_{\ell m} Y_{\ell m}({\hat\n}),
\end{equation}
with coefficients $a_{\ell m}$.  Polarization fluctuations (Stokes $Q$ and $U$) can be expanded in spin-2 spherical harmonics, $_2 Y_{\ell m}$, with $E$ and $B$ (grad- or curl-type) polarization coefficients
\begin{equation}
(Q \pm i U ) ({\hat\n}) = \sum_{\ell m} (a^{E}_{\ell m} \pm i a^{B}_{\ell m} ) _{\pm 2} Y_{\ell m}({\hat\n}).
\end{equation}

If the CMB is a Gaussian isotropic field, then the probability of a measurement of the sky given a model is described by a Multivariate Gaussian of the observed data:
\begin{equation}
L(\mathbf{d} | \mathbf{p}) = \frac{1}{2 \pi ^{N/2} | \mathbf{C}| ^{1/2} } \exp \left( -\frac{1}{2}  \mathbf{d} \, \mathbf{C}^{-1}  \mathbf{d}^{t} \right)
\label{pbLike}
\end{equation}
where $\mathbf{C}$ is the covariance of the data $\mathbf{d}$, and $\mathbf{p} $ is the set of model parameters. $\mathbf{C}
(\mathbf{p}) = \mathbf{S} (\mathbf{p}) + \mathbf{N}$, where $\mathbf{S}$ is the sky signal and $\mathbf{N}$ is the noise.  Since measurements of the sky are pixelated, the above likelihood is often called the "pixel-based likelihood" when estimated in the pixel domain.  It can be evaluated directly if the number of pixels is not too large, but for a full-sky experiment such as Planck with 5\arcm\ pixels it is impossible with current computers.    

\subsection{Exact likelihood in harmonic space}
\label{like-exact}

If the CMB is Gaussian, its statistical properties are represented fully by the underlying power spectrum $C_\ell$.  The multipole harmonic coefficients, $a^X_{\ell m}$ (where $X$ is $T$, $E$, or $B$), on different scales are independent of one another, and we can write
\begin{equation}
\left< (a^X_{\ell m})^* a^{X'}_{\ell' m'} \right >
 = \delta_{\ell \ell'} \delta_{m m'} C^{XX'}_{\ell}.
\label{eq:almX}
\end{equation}
The $a^X_{\ell m}$ are complex.  Under the assumption of Gaussianity, their real and imaginary parts are independent and Gaussian distributed, hence their phases are random. Since the CMB is real, they must satisfy $a^{X*}_{\ell m} = (-1)^m a^X_{\ell -m}$.  This means that $a^X_{\ell 0}$ is real.

For $m = 0$ we have
\begin{equation}
P(\bmath{a}_{\ell 0} | \bmath{C}_{\ell}) \mathrm{d}\bmath{a}_{\ell 0}
= \frac{1}{(2\pi)^{3/2} |\bmath{C}|^{1/2}} 
\exp \left \{ -\frac{1}{2}
\bmath{a}_{\ell 0}^T \bmath{C}_{\ell}^{-1}\bmath{a}_{\ell 0}
\right \}.
\end{equation}

For $m\ne0$,  we have for the real part of the $a^X_{\ell m}$
\begin{equation}
P(\Re\{\bmath{a}_{\ell m}\} | \bmath{C}_{\ell}) \mathrm{d}\Re\{\bmath{a}_{\ell m}\}
= \frac{1}{\pi^{3/2} |\bmath{C}|^{1/2}} 
\exp \left \{ -
\Re\{\bmath{a}_{\ell m}\} \bmath{C}_{\ell}^{-1}\Re\{\bmath{a}_{\ell m}\}
\right \},
\end{equation}
where
\begin{equation}
\bmath{a}_{\ell m} = 
\left(
\begin{array}{c}
{a}_{\ell m}^{T} \\
{a}_{\ell m}^{E} \\
{a}_{\ell m}^{B}
\end{array}
\right)
\end{equation}
and
\begin{equation}
\bmath{C}_{\ell} = 
\left(
\begin{array}{ccc}
{C}_{\ell}^{TT} & {C}_{\ell}^{TE} & 0 \\
{C}_{\ell}^{TE} & {C}_{\ell}^{EE} & 0 \\
0		  &       0           & {C}_{\ell}^{BB}
\end{array}
\right),
\end{equation}
Similarly for the imaginary part. 

Combining together all the values of $m$, we find, for a particular $\ell$:
\begin{equation}
P({\bmath{\hat C}}_\ell | \bmath{C}_\ell) \propto
| {\bmath{\hat C}}_\ell | ^{\frac{2\ell-3}{2}} 
| \bmath{C}_\ell | ^{-\frac{2\ell+1}{2}} 
\exp \left\{ 
-\frac{2 \ell + 1}{2} 
\mathrm{Tr} \left({\bmath{\hat C}_\ell} \bmath{C}_{\ell}^{-1} \right)
\right \}.
\end{equation}
where
\begin{equation}
{\hat C}_\ell^{XX'} = \sum_{m = -\ell}^{\ell}
\frac{(a^X_{\ell m})^* a_{\ell m}^{X'}}{2\ell+1},
\end{equation}
and the normalisation is independent of ${\bmath C}_\ell$ and ${\bmath{\hat C}}_\ell$.  In data analysis, the measured power spectrum ${\bmath{\hat C}}_\ell$ is a fixed quantity, hence the dependence of the likelihood on this value is generally dropped. In this case, up to a constant, we can write the log-likelihood as
\begin{equation}
\label{exactfullsky}
-2\ln P(\bmath{\hat C}_\ell | \bmath{C}_\ell)
 = (2\ell + 1) \left( 
\ln |\bmath{C_\ell}| + 
\mathrm{Tr} \left(
 {\bmath{\hat C}_\ell} \bmath{C}_\ell^{-1} \right)
\right),
\end{equation}
i.e., the inverse Wishart distribution.  If we consider only one measurement, e.g., one  $T$-mode or $B$-mode,  we can write the likelihood function as \citep {DABJK00}
\begin{equation}
- 2\ln P({\hat C}_\ell | C_\ell)
 = (2\ell + 1) \left( 
\ln \left (\frac{C_\ell}{{\hat C}_\ell} \right) + 
\frac {{\hat C}_\ell} {C_\ell}
\right),
\label{eq:single_like}
\end{equation}
i.e., the inverse Gamma distribution, where $C_\ell$ is the theoretical value of $C_\ell^{TT}$ (or $C_\ell^{BB}$) and ${\hat C}_\ell$ is the measured value.  

We can write an exact expression for the likelihood function for our measured power spectrum ${\hat C}_\ell$ given the true underlying power spectrum $C_\ell$, which is a function of cosmological parameters.  Since this likelihood is usually considered in the context of data analysis, it is common to regard the measured ${\hat C}_\ell$ as fixed quantities, and to write the likelihood as
\begin{equation}
\ln P({\hat C}| C)
 = \sum_\ell
-\frac{(2\ell + 1)}{2} \left( 
\ln \left (\frac{C_\ell}{{\hat C}_\ell} \right) + 
\frac {{\hat C}_\ell} {C_\ell}
\right) + \mathrm{constant,}
\end{equation}
where the constant depends on ${\hat C}_\ell$.  For fixed ${\hat C}_\ell$, this function peaks at $C_\ell = {\hat C}_\ell$.  However if we wish to consider the likelihood as a function of ${\hat C}_\ell$ then it is necessary to include the ${\hat C}_\ell$-dependence of the likelihood, in which case it should be written as
\begin{equation}
\ln P({\hat C}| C)
 = \sum_\ell
\frac{(2\ell - 1)}{2} \ln{\hat C}_\ell
- \frac{(2\ell + 1)}{2} 
\left( 
\ln C_\ell + 
\frac {{\hat C}_\ell} {C_\ell}
\right) + \mathrm{constant}.
\label{eq:exact}
\end{equation}
For a fixed underlying power spectrum $C_\ell$, this function peaks
at ${\hat C}_\ell = \left(\frac{2\ell-1}{2\ell+1}\right) C_\ell$.

This is adequate for a full-sky experiment with an infinitely narrow beam and no instrumental noise. For a partial or ``cut'' sky it is necessary to account for the correlations between the ${\hat C}_\ell$ that are introduced.  In addition,  real experiments always have non-uniform noise, and must estimate bandpowers  (${\hat C}_B$) rather than individual ${\hat C}_\ell$.  We need to find an appropriate likelihood function that includes the correct correlations and accounts properly for noise.

\subsection{Approximating the likelihood at high-$\ell$}
\label{like-approx}

Although the signal and the noise can be assumed Gaussian, the distribution of the band powers is non-Gaussian. This is the so called 'cosmic bias'---the distribution is skewed towards higher $C_{l}$ values. This effect is most noticeable at low ${\ell}$ where cosmic variance of the signal dominates the error bars.  As we will see, the full likelihood includes the cosmic bias whereas the Gaussian approximation of the Fisher matrix does not.  The Joint likelihood for temperature and polarization carries an extra complication in that one has to find an approximation that  properly accounts for the temperature and polarization cross power.  We will start by considering the current approximations derived for temperature alone, followed by an account on existing attempts to extend it to a joint likelihood for temperature and polarization.

\subsubsection{Temperature only}

{\bf Gaussian Likelihood}---The first level of approximation is to use a likelihood that is Gaussian in the ${\hat C}_\ell$ \citep {DABJK00}, i.e.,
\begin{equation}
P({\bmath{\hat C}}| \bmath{C}) \propto
\exp \left\{
-\frac{1}{2} ({\bmath{\hat C}} - \bmath{C})^T
\bmath{S}^{-1} ({\bmath{\hat C}} - \bmath{C})
\right\},
\end{equation}
where $\bmath{C}$ is a vector of $C_\ell$ values (and similarly
$\bmath{\hat C}$)
and $\bmath{S}^{-1}$ is the inverse signal covariance matrix.  However this likelihood function is well-known to be biased, (see e.g.,  \citet{DABJK00, smith06}).   This Gaussian likelihood can be implemented in two ways.  The version discussed in \citet{DABJK00} considers the case where the signal covariance matrix is derived from the measured power spectrum, rather than the theoretical power spectrum.  This results in an overestimation of the errors if the measured power spectrum has fluctuated upwards, and hence upward fluctuations are given less weight, leading to an overall downward bias.   The covariance matrix can instead be computed from the theoretical power spectrum, as used in \citet {DA-Ver03}, leading to an overestimation of the amplitude.

\medskip\noindent
{\bf Offset Lognormal Likelihood}---A better approximation to the likelihood is the Offset Lognormal approximation \citep {DABJK00}, given by

\begin{equation}
P_{LN}({\bmath{\hat C}}| \bmath{C}) \propto
\exp \left\{
-\frac{1}{2} ({\bmath{\hat z}} - \bmath{z})^T
\bmath{M}({\bmath{\hat z}} - \bmath{z})
\right\},
\end{equation}
where $z_\ell = \ln(C_\ell + x_\ell)$ and
the matrix $\bmath{M}$ is related to the inverse
covariance matrix by
\begin{equation}
M_{\ell \ell'} = 
(C_\ell + x_\ell) S^{-1}_{\ell \ell'} (C_\ell' + x_\ell')
\end{equation}
(The offset factors $x_\ell$ are simply a function of the
noise and beam of the experiment.)
This likelihood function is still slightly biased, but in the opposite
direction to that of the Gaussian likelihood.

To some extent the Offset Lognormal approximation addresses this issue, in that the transformation of variables from $C_{\ell} $ to $Z_{\ell} = \ln(C_{\ell} + x_{\ell}) $ has a constant curvature matrix.  Hence the uncertainties on the $Z_{\ell}$ do not depend on the $Z_{\ell,{\rm estimated}}$, avoiding the cosmic bias. To proceed, assume a normal distribution in this new variable $Z_{\ell}$ instead of $C_{\ell}$. The likelihood is now closer to the exact one. Nevertheless, there is still a small bias opposite to that introduced by the Gaussian approximation in $C_{\ell}$.

\medskip\noindent
{\bf WMAP Likelihood}---Taking advantage of the fact that the bias for the Offset Lognormal likelihood is opposite to that introduced by the Gaussian likelihood, the  WMAP team defined a likelihood that is a weighted combination of the two \citep {DA-Ver03}:

\begin{equation}
\ln P_\mathrm{WMAP}({\bmath{\hat C}}| \bmath{C}) = 
\frac{1}{3}\ln P_{\mathrm{Gauss}}({\bmath{\hat C}}| \bmath{C})
+ \frac{2}{3}\ln P_{\mathrm{LN}}({\bmath{\hat C}}| \bmath{C})
\end{equation}
This likelihood is a significantly better approximation for the case of a Gaussian \hbox{CMB}.  However, the fact that a likelihood function is accurate in the absence of correlations (when the probability of a measured power spectrum is purely a function of the input power spectrum rather than having any additional dependence on the cosmology) is not a guarantee that it will perform well when applied to a non-Gaussian sky. For instance, \citet{smith06} have shown that the WMAP likelihood gives significant biases in the dark energy parameter, $w$, when considering the lensed B-mode power spectrum.

\medskip\noindent
{\bf Equal Variance Likelihood}---The equal variance likelihood proposed by \citet{DABJK00} is given by:
\begin{equation}
\ln{L} = -\frac{1}{2}G \left[  e^{-(z-{\hat z} ) }- \left(1-(z-{\hat z}) \right)    \right]
\end{equation}
with
\begin{equation}
z= \ln \left( q_{b} + q_{b}^{N} \right)
\end{equation}
and

\begin{equation}
G = \left [   e^{-\sigma_{z}} - (1-\sigma_{z})  \right]^{-1} , \sigma_{z} =\frac{\sqrt{{\cal{F}}_{bb'}^{-1}}}{(q_{b} + q_{b}^{N})}       
\end{equation}
The noise offset $q_{b}^{N}$ is estimated using the equation of the maximum likelihood solution for the $q_{b}$, replacing the observed map with the average of the noise Monte Carlo simulation power spectra $ \left<  \tilde{N_{\ell}} \right > $.

\medskip\noindent
{\bf SCR Likelihood}---\citet{smith06} developed a new likelihood to tackle the non-Gaussianity of the lensed sky for studies of the $B$-mode power spectrum; however, this likelihood has not been extended to the joint probability distribution for all modes. By considering the curvature (with respect to the measured ${\hat C}_\ell$) of the exact log-likelihood expression for a Gaussian sky (equation \ref{eq:exact}) at its peak, and also the third derivative, a new likelihood can be derived which is Gaussian in $x_\ell = ({\hat C}_\ell)^{\frac{1}{3}}$, where both the second and third derivatives with respect to $x_\ell$ take the correct values at the peak of the likelihood.

The SCR likelihood approximates the normalised distribution $P({\hat C}_\ell | \bmath{\theta})$ as Gaussian in some function of the ${\hat C}_\ell$, and takes the form
\begin{equation}
\ln P({\hat C}_\ell|\bmath{\theta}) \approx \ln A -\frac{1}{2}
\sum_{\ell \ell'} M^{-1}_{\ell \ell'} ({\hat x}_\ell -\mu_\ell)
({\hat x}_{\ell'} -\mu_{\ell'}),
\label{eq:new-like}
\end{equation}
where 
\begin{eqnarray}
{\hat x}_\ell &=& {\hat C}_\ell^{1/3} \\
\mu_\ell &=& \left(\frac{2\ell-1}{2\ell+1}C_\ell \right)^{1/3},
\end{eqnarray}
and
\begin{equation}
M^{-1}_{\ell \ell'} = 3 C_\ell^{2/3} \left( \frac{2\ell-1}{2\ell+1}\right)^{1/6}
S^{-1}_{\ell \ell'}  \; 
3 C_{\ell'}^{2/3} \left( \frac{2\ell'-1}{2\ell'+1}\right)^{1/6}.
\end{equation}
Here $S_{\ell \ell'}$ is the covariance matrix of the measured ${\hat C}_\ell$ at parameters $\bmath\theta$. The normalisation
is
\begin{equation}
A^{-1} \propto ({{\rm det} M_{\ell \ell'}})^{1/2} \prod_\ell \mu_\ell^2,
\end{equation}
which can be approximated by $A \propto \prod_{\ell} 1/C_\ell$.

Applying this expression to a Gaussian simulation gives results almost indistinguishable from the exact likelihood expression, as shown in \citet{smith06}.
 Ignoring the  $(2\ell-1)/(2\ell+1)$ factors, however, is a bad approximation at low $\ell$, since it ignores the fact that, for a fixed $C_\ell$, the peak of the likelihood is slightly below $C_\ell$. This ends up translating to an underestimation of the theoretical power spectrum.  The performance of the Gaussian, WMAP, and two versions of the SCR likelihoods on full-sky lensed simulations was compared in \citet{smith06}.  The WMAP likelihood function gives very different results to the new SCR likelihood function, and shows a significant bias in the dark energy parameter, $w$.

\subsubsection{Temperature + polarization}
\label{like-pol}

We cannot merely extend the above approximations to build a joint likelihood for temperature and polarization.   For instance, assume that we approximate the likelihood for $TE$ as a Gaussian. The mode and variance of the Gaussian distribution for $TE$ depend on $TT$ and $EE$. Hence it is not enough to consider the joint likelihood as a product of independent $TT$, $EE$, and $TE$ likelihoods. Indeed the trick is to find a way of coupling these components reliably.  Furthermore, given that the $TE$ power spectrum is at times negative, we cannot build a Joint likelihood as a product of independent Offset Lognormal likelihoods. As a quick fix, one could try the following:

\medskip\noindent
{\bf Offset Lognormal Bandpower Likelihood}---This likelihood is a joint likelihood for temperature and polarization built as a Gaussian for $TE$ and Offset Lognormal for $TT$, $EE$, and $BB$.  However, this approximation does not properly take into account temperature and polarization cross power. 

\medskip

The following two likelihoods, the Hamimeche and Lewis and {\tt XFaster} likelihoods, do attempt to take temperature and polarization cross power into account properly.

\medskip\noindent
{\bf Hamimeche and Lewis (HL) Likelihood}---The \citet{HL_like} likelihood generalizes the full-sky exact likelihood given by Equation~\ref{exactfullsky} to the cut sky by considering a quadratic expression of the form
\begin{equation}
\ln{L}(C_{\ell} | {\hat C}_{\ell}) = - \frac{1}{2} \frac{2\ell+1}{2} \sum_{i} \left[ g(D_{\ell,ii} \right]^{2} = \frac{2\ell+1}{2} {\rm Tr} \left[ \mathbf{g}(\mathbf{D}_{\ell})^{2} \right],  
\end{equation}
where $g(x)= {\rm sgn}(x-1) \sqrt{2(x- \ln{x} -1)}$
and $\left[\mathbf{g}(\mathbf {D}_{\ell})\right]_{ij} =  g(D_{\ell,ii})\delta_{ij}$.
This approximation involves a fiducial model so that the covariance can be pre-computed. It is assumed that the matrix of estimators $\mathbf{{\hat C}}_{\ell}$ is positive/definite, although this assumption may break down for some estimators at low-$\ell$.

\medskip\noindent
{\bf {\tt XFaster}  Likelihood}---The {\tt XFaster} likelihood, introduced in \citet{XFaster-like1} and \citet{XFaster-like2}, takes the following form for temperature alone:
\begin{equation}
\ln{L}= -\frac{1}{2} \sum_{\ell} g (2 \ell +1) \left( \frac{\tilde{C_{\ell}^{obs}}}{ \left( \tilde{C_{\ell}} + \left< \tilde{N_{\ell}}\right> \right)} + \ln \left( \tilde{C_{\ell}} + \left< \tilde{N_{\ell}}\right> \right) \right),
\label{XFaster-likex}
\end{equation}
where a tilde connotes a quantity estimated on the cut-sky.  The power spectrum is parameterized through a set of deviations $q_{\ell}$ from a template full-sky  spectrum $C_{\ell}^{(S)}$,
\begin{equation}
\tilde{C}_{\ell} = \sum_{\ell'} K_{\ell \ell'} B_{\ell'}^{2} F_{\ell'} C_{\ell'}^{(S)} q_{\ell'},
\label{Cl-cut}
\end{equation}
where $K_{\ell \ell'} $ is the coupling matrix due to the cut sky observations, $ B_{\ell}$ expresses the effect of a finite beam, and 
$F_{\ell}$ is a transfer or filter function accounting for the effect of pre-filtering the data in both time and spatial domains.

Extending to polarization, we have
\begin{equation}
\ln{L}= -\frac{1}{2} \sum_{\ell} g (2 \ell +1) \left(Tr \left ({\tilde{\mathbf{D}_{\ell}^{obs}}}{ \left( \tilde{\mathbf{D}_{\ell}} + \left< \tilde{\mathbf{N}_{\ell}}\right> \right)^{-1}} \right)+ ln  \left| \tilde{\mathbf{D}_{\ell}} + \left< \tilde{\mathbf{N}_{\ell}}\right> \right| \right),
\label{XFaster-likey}
\end{equation}
where the matrix $\mathbf{C}$ is block diagonal: $\tilde{\mathbf{C}} \rightarrow diag(
\tilde{\mathbf{D}}_{\ell_{min}},
\tilde{\mathbf{D}}_{\ell_{min}+1},\dots,
\tilde{\mathbf{D}}_{\ell_{max}}) $, with each multipole's covariance
given by the $3 \times 3$ matrix
\begin{equation}
\tilde{\mathbf{D}}_{\ell} = 
\left(
\begin{array}{ccc}
\tilde{C}_{\ell}^{TT} & \tilde{C}_{\ell}^{TE} & \tilde{C}_{\ell}^{TB} \\
\tilde{C}_{\ell}^{TE} & \tilde{C}_{\ell}^{EE} & \tilde{C}_{\ell}^{EB}  \\
\tilde{C}_{\ell}^{TB} &   \tilde{C}_{\ell}^{EB} & \tilde{C}_{\ell}^{BB} \\
\end{array}
\right).
\end{equation}
This likelihood follows intuitively from the full-sky exact likelihood, the Inverse Wishart distribution, as given by Equation~\ref{exactfullsky}.

\subsection{The likelihood at low-$\ell$}
\label{like-lowl}

The pixel-based Multivariate Gaussian likelihood given by Equation~\ref{pbLike} can be computed up to $\ell \simeq 30, 40$, and is adequate for comparison with {\tt Xfaster} as shown in \citep{XFaster-like2}.  (Faster methods (see summary by \citet{varenna10}) have been developed, but are not necessary here.)  We use an implementation known as {\tt Bflike}, described as follows.

A CMB map can be written as an ordered vector ${\bf d} =
(T_{i_1},T_{i_2},...,T_{n_T},Q_{j_1},Q_{j_2},...Q_{n_P},U_{j_1},U_{j_2},...U_{n_P})$,
comprising all pixels with valid observations. In general $n_T \ne n_P $ and the sets of indexes of
temperature and polarization measurements will be different. Assuming
that both CMB and noise fluctuations in each pixel
are Gaussian-distributed with zero mean, the likelihood for ${\bf d}$
has the functional form given in equation (1), where the covariance
matrix has a block structure:
\begin{equation}
\label{eq:TQUcovar}
{\bf C} = 
\begin{pmatrix}
<TT>_{(n_T \times n_T)} & <TQ>_{(n_T \times n_P)} & <TU>_{(n_P \times n_P)} \\
<QT>_{(n_P \times n_T)} & <QQ>_{(n_P \times n_P)} & <QU>_{(n_P \times n_P)} \\
<UT>_{(n_P \times n_T)} & <UQ>_{(n_P \times n_P)} & <UU>_{(n_P \times n_P)} 
\end{pmatrix}
\end{equation}
Correlations between, e.g., temperature measurements in pixels $i_1$ and $i_2$
can be written as:
\begin{equation} 
\label{eq:ttcorr}
\langle T_{i_1} T_{i_2} \rangle =
  \sum_{\ell=2}^{\ell_{\rm max}} {{2\ell +1}\over{4\pi}} {\hat C}_\ell P_\ell(\theta_{i_1i_2}) + {\bf N_{i_1i_2}},
\end{equation}
$P_\ell(x)$ are the ordinary Legendre polynomials, and $\theta_{i_1i_2}$ is the angle between the centers of pixels $i_1$ and $i_2$. Notice that the $\{C_\ell\}$ include the contribution of the beam and pixel window, i.e., $\{{\hat C}_\ell\} =\{C_\ell^{\rm th}\} b_\ell^2 w_\ell^2$, where $\{ C_\ell^{\rm th} \}$ is the theory power spectrum
and $b_\ell$ and $w_\ell$ are the harmonic transform of the beam and window functions respectively. For uncorrelated noise, $N_{ij} = n_i^2 \delta_{ij}$, where $n_i$ is the rms noise in pixel $i$.  In general, ${\bf N}$ is a dense matrix. Similar expressions hold for
correlations involving $Q$ and $U$ (see e.g., \citet{tegmarkcosta01}). The choice of $\ell_{\rm max}$ in Equation~\ref{eq:ttcorr} depends on several factors, including the
smoothing beam, the signal-to-noise ratio, and the pixelization scheme.


\section{Comparing likelihoods}
\label{like-comp}

\subsection{Simulations}
\label{sims}

To compare the performance of {\tt XFaster} with other likelihood functions, we use simulations developed within the Planck CTP working group.  Planck \citep{bluebook06, planckmission09} is a full-sky experiment covering frequencies from 30 to 857\,GHz with beams ranging in size from 33\arcm\ to 5\arcm.  A full description of the  simulations is given in \citet{varenna10,XFaster-like2}.  
Practical considerations of computational resources having to do with the size of the time-ordered data (TOD), the number of pixels in the maps, and the number of multipoles that had to be calculated, dictated the choice of the 70\,GHz channel for the simulations. Higher frequency channels have higher angular resolution and sensitivity, and will extend to smaller angular scales with reduced error bars.  Recent increases in computational capability make it possible now to generate thousands of Monte Carlo simulations at the higher frequencies as well.  Results will be presented in a future publication \citet{GRhfiXFaster10,varenna10}.

The 70\,GHz simulations used here include the CMB, realistic detector noise, and noise induced by temperature fluctuations of the 20-K hydrogen sorption cooler.  To mimic the sensitivity of combination of channels, as would be used for separation of foregrounds with real data, the white noise level was taken to be lower than that expected for the 70\,GHz channel alone.  The white noise per sample was 2025.8\,$\mu$K and the $1/f$ noise power spectrum had knee frequency 0.05\,Hz and slope $-1.7$.  The input sky signal used to generate the ``observed map'' was the CMB map derived from the Planck CMB reference sky available in\footnote{http://www.sissa.it/~planck/reference-sky/CMB/alms/alm-cmb-reference-template-microKthermodynamic-nside2048.fits}
which uses the WMAP 1-year $a_{lm}$ up to $\ell = 70$ to generate the \hbox{CMB}. In other words, the large scale structure of the observed map is a WMAP-constrained realization.

The TOD were generated using modules of the Planck simulator pipeline, {\tt LevelS} \citep{levels}.  Maps were made from the simulated time-ordered data (TOD) using the destriping code Springtide \citep{cambridge, helsinki, paris, trieste, springtide}.  Where a sky cut was applied, it was made at the boundary where the total intensity of the diffuse foregrounds and point sources exceeded twice the CMB sigma.  Pixels missing due to the scanning strategy were masked.
The beams of the detectors have FWHMs of 13\arcm--14\arcm, so the maps were made with $N_{\mathrm{side}} = 1024$, corresponding to a pixel size of 3\parcm4. Two sets of maps were provided, one 12-detector map to be used in the auto-spectrum mode,  and three 4-detector maps to be used in the cross-spectrum mode.  

Two cases were considered.  The first, called Phase~2a for historical reasons, assumed symmetric Gaussian beams with FWHM of 14\arcm.  The second, Phase~2b, assumed elliptical Gaussians fit to the central parts of realistic beams calculated by a full diffraction code for the Planck optical system.  

One hundred Monte Carlo signal simulations were generated from the best fit WMAP + CBI + ACBAR $\Lambda$CDM power spectrum\footnote{available in http://lambda.gsfc.nasa.gov/product/map/dr1/lcdm.cfm}, with BB mode power set to zero.  For the symmetric beam case the signal could be simulated in the map domain, while noise was generated in the time domain.  For the asymmetric beam case, both signal and noise simulations were generated in the time domain.

The ``low-$\ell$ dataset'' of the Phase2 simulations was generated directly at $N_{\rm side} =16$.  The procedure adopted ensured consistency between the low- and high-$\ell$ datasets used to test the {\tt XFaster} power spectrum and likelihood estimator; however, the low-$\ell$ dataset lacks the artifacts connected to smoothing and degradation of higher resolution maps that will be present in the final Planck maps.

As pointed out in \citet{XFaster-like2}, since the large scale structure of the observed map is derived from real observations, i.e., a WMAP constrained realization, it is not necessarily consistent with the best-fit spectrum at low multipoles.  This discrepancy is evident when comparing the cosmological parameters estimated with {\tt XFaster} power spectrum and likelihood and the Offset Lognormal likelihood with the theoretical best fit parameters. The Monte Carlo simulations, however, are realizations of the first year WMAP+CBI+ACBAR best fit $\Lambda$CDM power spectrum for Phases 2a and 2b, so such discrepancy is no longer present.  Parameters estimated from these Monte Carlo simulations maps are shown to be close to the WMAP best fit parameters.


\subsection{Comparisons}
\label{like-par}

\citet{XFaster-like2} (Figs. 14 and 15) showed that for $TT$ and $TE$, all of the likelihood approximations described above except the Gaussian likelihood converge to the same form at multipoles above 10 or 20.  For $EE$, however, the Gaussian and the Lognormal likelihood differs from the rest up to high $\ell \simeq10$.  \citet{XFaster-like2} (Fig. 17) also showed by comparing {\tt XFaster} with the pixel-based likelihood code ({\tt BFlike}) that a transition from low-$\ell$ to high-$\ell$ codes was appropriate for Planck in the range $\ell_{\rm trans} = 30$--$40$

Here we assess the performance of the {\tt XFaster} likelihood by comparing cosmological parameters obtained with the {\tt XFaster} and the Offset Lognormal Bandpower (i.e., Offset Lognormal likelihood for $TT$, $EE$, $BB$, and Gaussian for $TE$) likelihoods.  {\tt XFaster} computes the likelihood of a model passed to it by a modified version of the publicly available \texttt{CosmoMC} code\footnote {http://cosmologist.info/cosmomc/} (\citet{Lewis:2002ah})) for cosmological parameter Markov Chain Monte Carlo estimations.  There is no need for window functions or the band power spectrum itself.  The inputs are the raw pseudo-$C_{\ell}$ of the observations plus the kernel and transfer function required by {\tt XFaster} to relate the cut-sky pseudo-$C_{\ell}$ to the full-sky $C_{\ell}$. 

For the Offset Lognormal Bandpower likelihood, window functions are required that properly account for band power spectrum correlations.  We used two different window functions, top hat (box) and Fisher-weighted \citep{XFaster-like2}, written ${\cal{F}}_{bb}$.  For the sake of completeness we describe here how we derive ${\cal{F}}_{bb}$. To construct ${\cal{F}}_{bb}$ (following \citet{DABJK00} and \citet{XFaster-like2}) we define a logarithmic integral,
\begin{equation}
\mathcal{I}[f_{\ell}] = \sum_{\ell}  \frac{\ell + \frac{1}{2}}{\ell (\ell + 1)} f_{\ell},
\end{equation}
which is used to calculate the expectation values for the deviations $q_{b}$(when a shape model, $C_{\ell}^{S}$ is considered), or bandpowers $C_{b}$ (when $C_{\ell}^{S}$ is assumed to be flat).
\begin{equation}
\langle q_{b} \rangle = \frac{\mathcal{I}\left[ W_{\ell}^{b} \mathcal{C}_{\ell} \right] }{\mathcal{I} \left[ W_{\ell}^{b} \mathcal{C}_{\ell}^{(S)} \right] } \,\
\langle C_{b} \rangle = \frac{\mathcal{I} \left[ W_{\ell}^{b} \mathcal{C}_{\ell} \right] }{\mathcal{I} \left[ W_{\ell}^{b} \right] } .
\end{equation}
where $W_{\ell}^{b}$ is the band power window function, with $\mathcal{C}^{(S)} = \ell(\ell + 1) C_{\ell}^{(S)} / 2 \pi$.  We define normalized window functions 
\begin{equation}
\mathcal{I} \left [ W_{\ell}^{b} \mathcal{C}_{\ell}^{(S)} \right ] = 1,
\end{equation}
using the fact that
\begin{equation}
\langle ( \tC_{\ell}^{obs} - \tN_{\ell} ) \rangle \rightarrow \tC_{\ell}
\end{equation} 
to obtain
\begin{equation}
W_{\ell}^{b} = \frac{4 \pi}{(2 \ell + 1) } \sum_{b'} {\cal{F}}_{b b'}^{-1} \sum_{\ell'} g  (2 \ell' +1 ) \frac{\tC_{b' \ell'}^{(S)}}{(\tC_{\ell'}^{T} + \langle \tN_{\ell'}\rangle ) ^{2}} K_{\ell \ell'} F_{\ell} B_{\ell}^{2}.
\end{equation}
Extending to polarization:
\begin{equation}
W_{\ell}^{b} =  \frac{4 \pi}{(2 \ell + 1) } \sum_{b'} {\cal{F}}_{b b'}^{-1} \sum_{\ell'} g (2 \ell' +1 ) {\rm Tr}  \left[ \vW_{b' \ell'}  \vK_{\ell'} \right]
\end{equation}
where $\vW_{b \ell}  = { \tilde{\mathbf{D}}_{\ell}}^{-1}  \frac{\partial \mathbf{\tilde{S}}}{\partial q_{b}} {\tilde{\mathbf{D}}_{\ell}}^{-1}$, and $\vK_{\ell}$ gives the cut-sky response to the individual full-sky multipoles,
\begin{equation}
\vK_{\ell}= 
\left(
\begin{array}{ccc}
K_{\ell' \ell} F_{\ell}^{TT} B_{\ell}^2 &  _{\times}  K_{\ell' \ell} F_{\ell}^{TE} B_{\ell}^2 &  _{\times}  K_{\ell' \ell} F_{\ell}^{TB} B_{\ell}^2 \\
_{\times}  K_{\ell' \ell} F_{\ell}^{TE} B_{\ell}^2 &   _{+} K_{\ell' \ell} F_{\ell}^{EE} B_{\ell}^2 + _{-} K_{\ell' \ell} F_{\ell}^{BB} B_{\ell}^2 & \left( _{+} K_{\ell' \ell} - _{-} K_{\ell' \ell} \right) F_{\ell}^{EB} B_{\ell}^2 \\
_{\times}  K_{\ell' \ell} F_{\ell}^{TB} B_{\ell}^2 &   \left( _{+} K_{\ell' \ell} - _{-} K_{\ell' \ell} \right) F_{\ell}^{EB} B_{\ell}^2 &  _{+} K_{\ell' \ell} F_{\ell}^{BB} B_{\ell}^2 + _{-} K_{\ell' \ell} F_{\ell}^{EE} B_{\ell}^2 \\
\end{array}
\right).
\end{equation}

The parameters calculated are: baryonic, cold dark matter, and cosmological constant densities, $\omega_b=\Omega_bh^2$, $\omega_c=\Omega_ch^2$, and $\omega_{\Lambda}=\Omega_{\Lambda} h^2$, respectively; the ratio of the sound horizon to the angular diameter distance at decoupling, $\theta_s$; the spectral index of the initial fluctuation spectrum, $n_s$; the overall normalization of the spectrum $\log[10^{10} A]$ at $k=0.05$ Mpc$^{-1}$ ($A_s$); the Hubble constant $H_{0}$; and the reionization redshift $z_{re}$.  We treat $\tau$ in two different ways.  \citet{XFaster-like2} showed that ``high $\ell$'' codes could be used to determine parameters from the ``observed map'' quite well if $\tau$ was fixed in the fit to the value of the input model.  This works because $\tau$ is constrained primarily by data at $\ell<30$.  We indicate when $\tau$ is held fixed.

\begin{figure}
\begin{center}
\vbox{\hglue -3pt
\includegraphics[width=17.9cm]{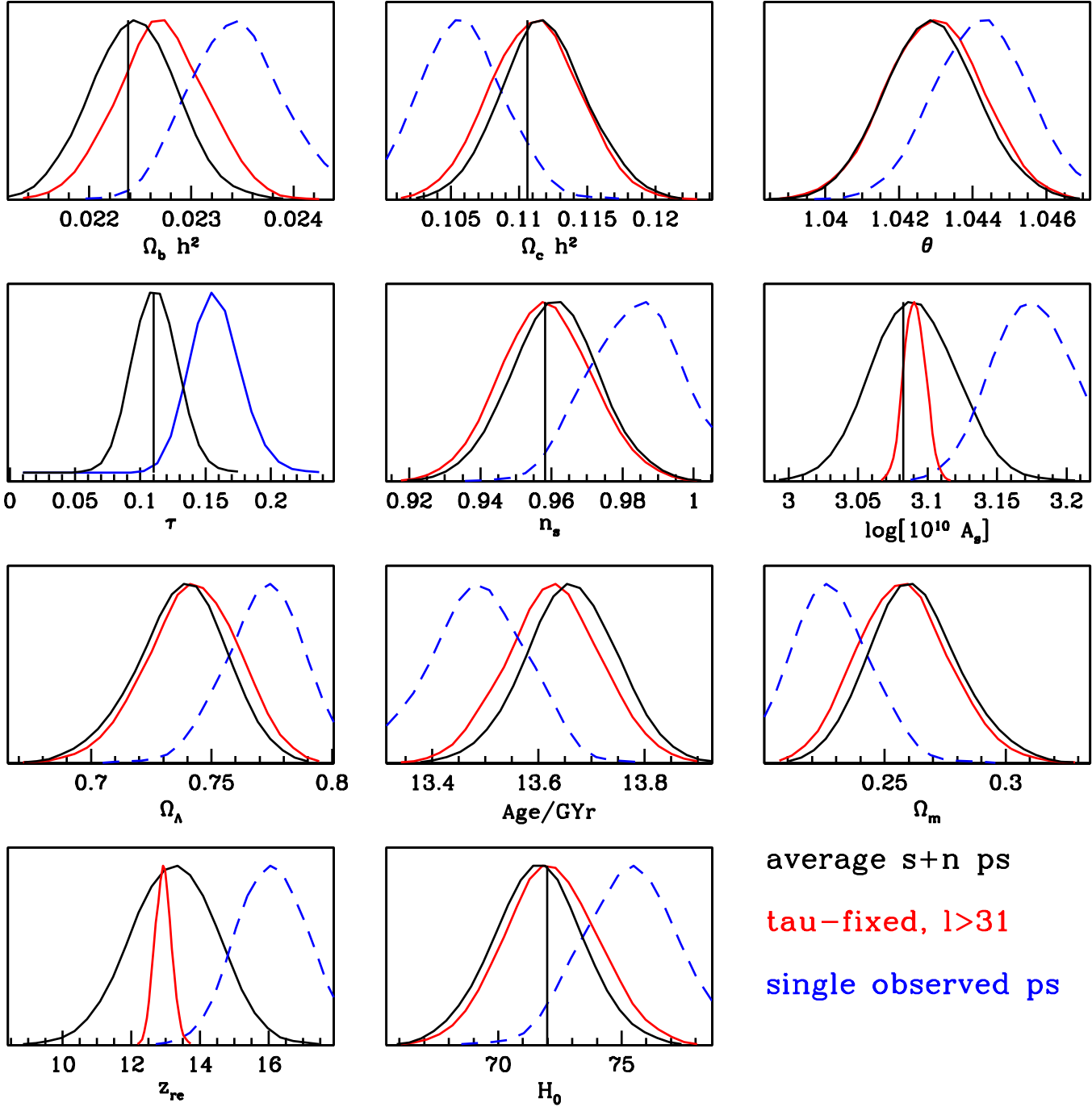}
}
\caption{Marginalised parameter distributions from {\tt Xfaster}, for the observed map (blue dashed lines), the ensemble average of Monte Carlo simulations (black solid lines), and for the observed map holding $\tau$ fixed at its input value and using the {\tt Xfaster} likelihood only for $\ell>30$ (solid red lines).   Input parameter values are marked by vertical lines.  The input parameters are recovered quite well from the ensemble average (see text), and also for the observed map when $\tau$ is fixed and low multipoles are not included in the parameter fits.  Figure reproduced from \citet{XFaster-like2}}
\label{par-XFaster-mc-obs-fixtau}
\end{center}
\end{figure}

Figure~\ref{par-XFaster-mc-obs-fixtau}, from \citet{XFaster-like2}, shows  one-dimensional marginalised parameter distributions from {\tt Xfaster}, for three cases: 1)~the observed map; the ensemble average of Monte Carlo simulations; and the observed map, but holding $\tau$ fixed at its input value and using the {\tt Xfaster} likelihood only for $\ell>30$.  The input parameters are recovered quite well from the ensemble average.  The red lines show that if $\tau$ is fixed, high-$\ell$ codes such as {\tt XFaster} can ignore the low multipoles that they are not designed to calculate, and still recover the other input parameters quite well. 

\begin{figure}
\begin{center}
\vbox{\hglue -3pt
\includegraphics[width=17.8cm]{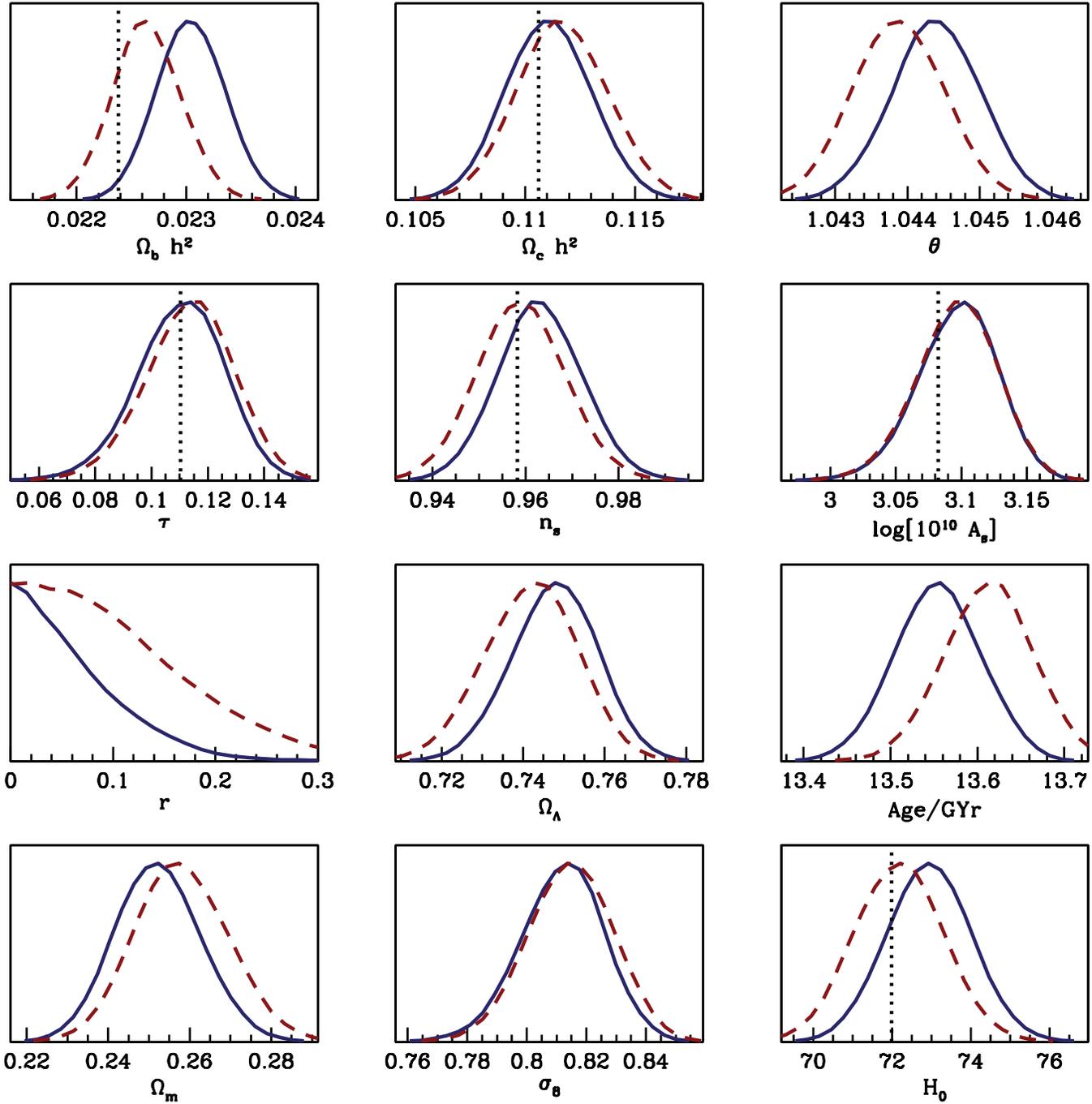}
}
\caption{Marginalised parameter distributions from the Offset Lognormal Bandpower likelihood (Offset Lognormal for $TT$, $EE$, $BB$, Gaussian for $TE$), for a Fisher-weighted (${{\cal{F}}_{bb}}$) window function.  Blue solid lines are from the observed map.  Red dashed lines are from the ensemble average of Monte Carlo simulations.  Input parameter values are marked by vertical lines.}
\label{par-Offlogn-obs-avg}
\end{center}
\end{figure}

Figure~\ref{par-Offlogn-obs-avg} shows parameter distributions from the Offset Lognormal Bandpower likelihood for a Fisher-weighted (${{\cal{F}}_{bb}}$) window function, for both the observed map and for the ensemble average of Monte Carlo simulations.  The input parameters are recovered from the ensemble average simulated data, but not from the observed map, particularly $A_s$.  This is a not a surprise.  As described in \S\,3.1, the observed map is fixed by WMAP for $2\le\ell\le70$.  It is therefore affected by low-$\ell$ anomalies arising from any cause, including the details of WMAP processing. The WMAP best-fit parameters, on the other hand, are obtained with heavy marginalization of the low-$\ell$ points by foregrounds, and are therefore little affected by the low-$\ell$ anomalies.   Since the Monte Carlo simulations are realizations of the WMAP best-fit model parameters, we expect no systematic bias from the ensemble of simulations, but significant offsets from the parameters derived from the observed map, as confirmed.

\begin{figure}
\begin{center}
\vbox{
\includegraphics[width=17.9cm]{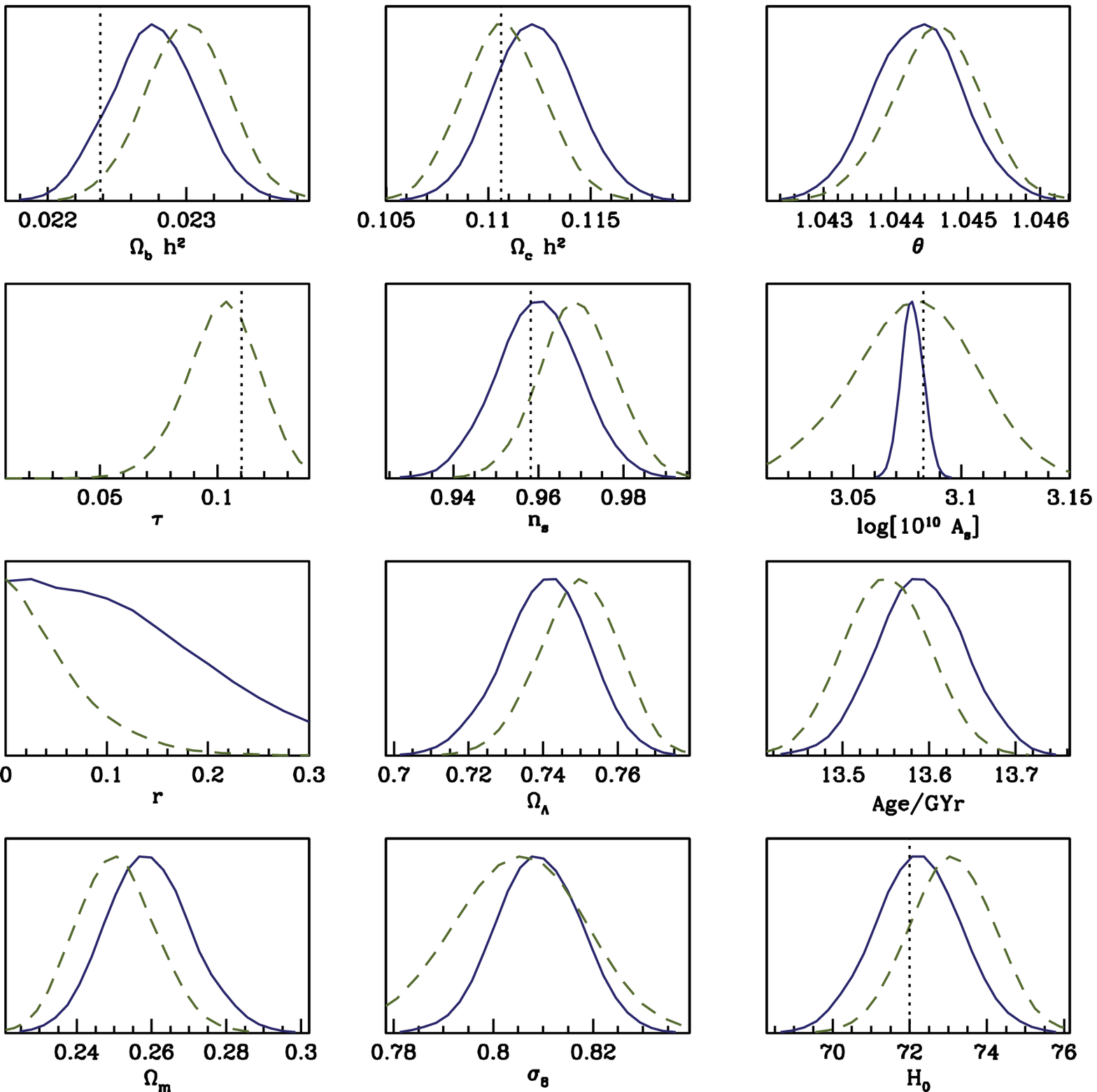}
}
\caption{Marginalised parameter distributions from the Offset Lognormal Bandpower likelihood for a top-hat window function, for the observed map.  Green dashed lines are computed from all multipoles, with $\tau$ free to vary, while for the blue solid lines use only $\ell>30$, with $\tau$ fixed at the input value.  Input parameter values are marked by vertical lines.}
\label{par-XOfflogn-obs-tau}
\end{center}
\end{figure}

Figure~\ref{par-XOfflogn-obs-tau} shows the effect of fixing $\tau$ in the case of the Offset Lognormal Bandpower likelihood for a top-hat window function for the observed map.  Green dashed lines are computed from all multipoles, with $\tau$ free to vary, while for the blue solid lines use only $\ell>30$, with $\tau$ fixed at the input value.  As with {\tt XFaster} in Figure~\ref{par-XFaster-mc-obs-fixtau}, fixing $\tau$ and ignoring low multipoles gives good results with the other parameters.

\begin{figure}
\begin{center}
\hglue -11pt
\hbox{
\includegraphics[width=18.05cm]{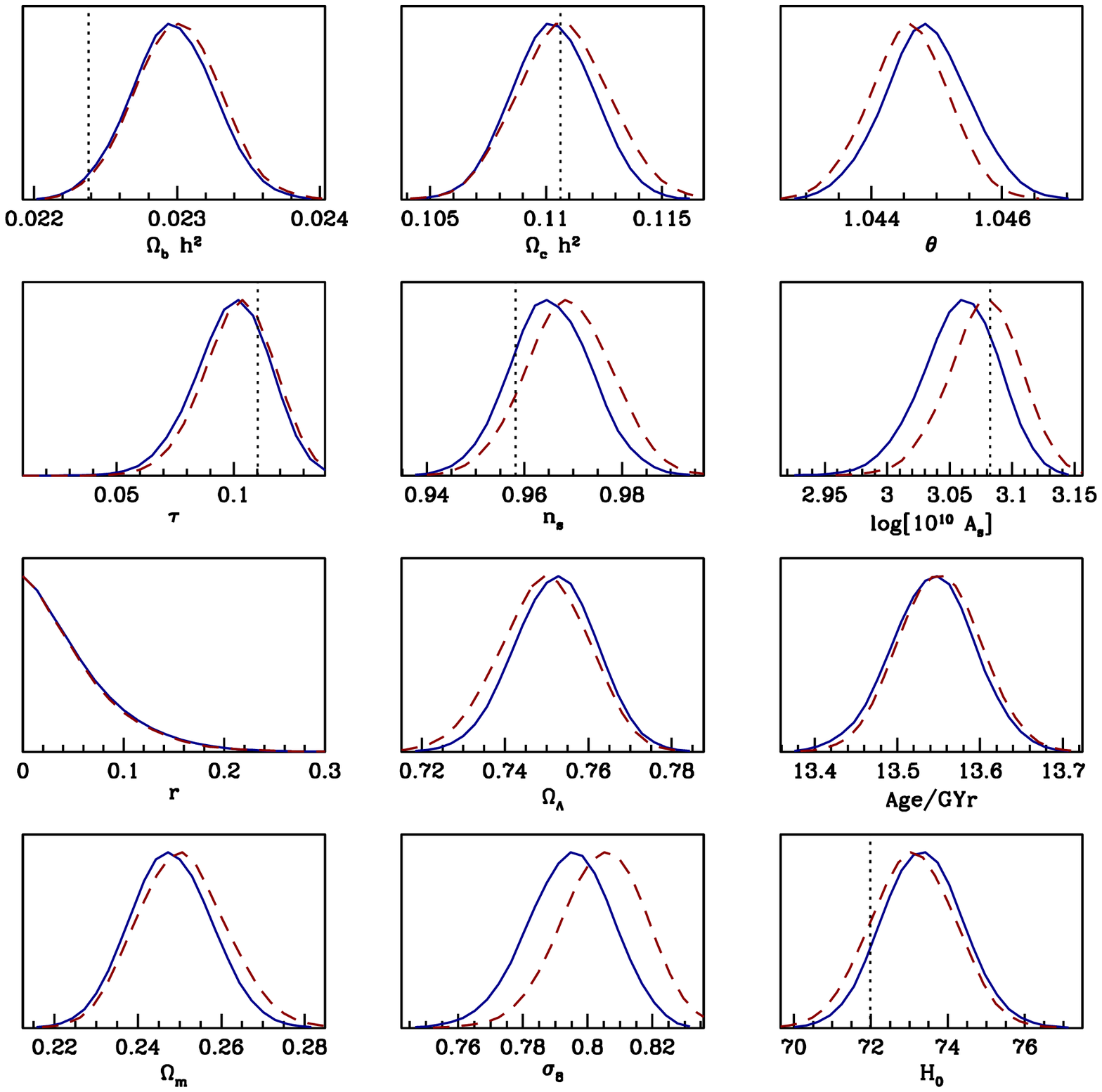}
}
\caption{Comparison of symmetric (red dashed lines) and asymmetric (blue solid lines) beams in marginalised parameter distributions from the Offset Lognormal Bandpower likelihood, for the observed map.  $\tau$ is not fixed.} 
\label{par-Offlogn-symm-asymm-obs}
\end{center}
\end{figure}

\begin{figure}
\hglue -3pt
\hbox{
\includegraphics[width=17.8cm]{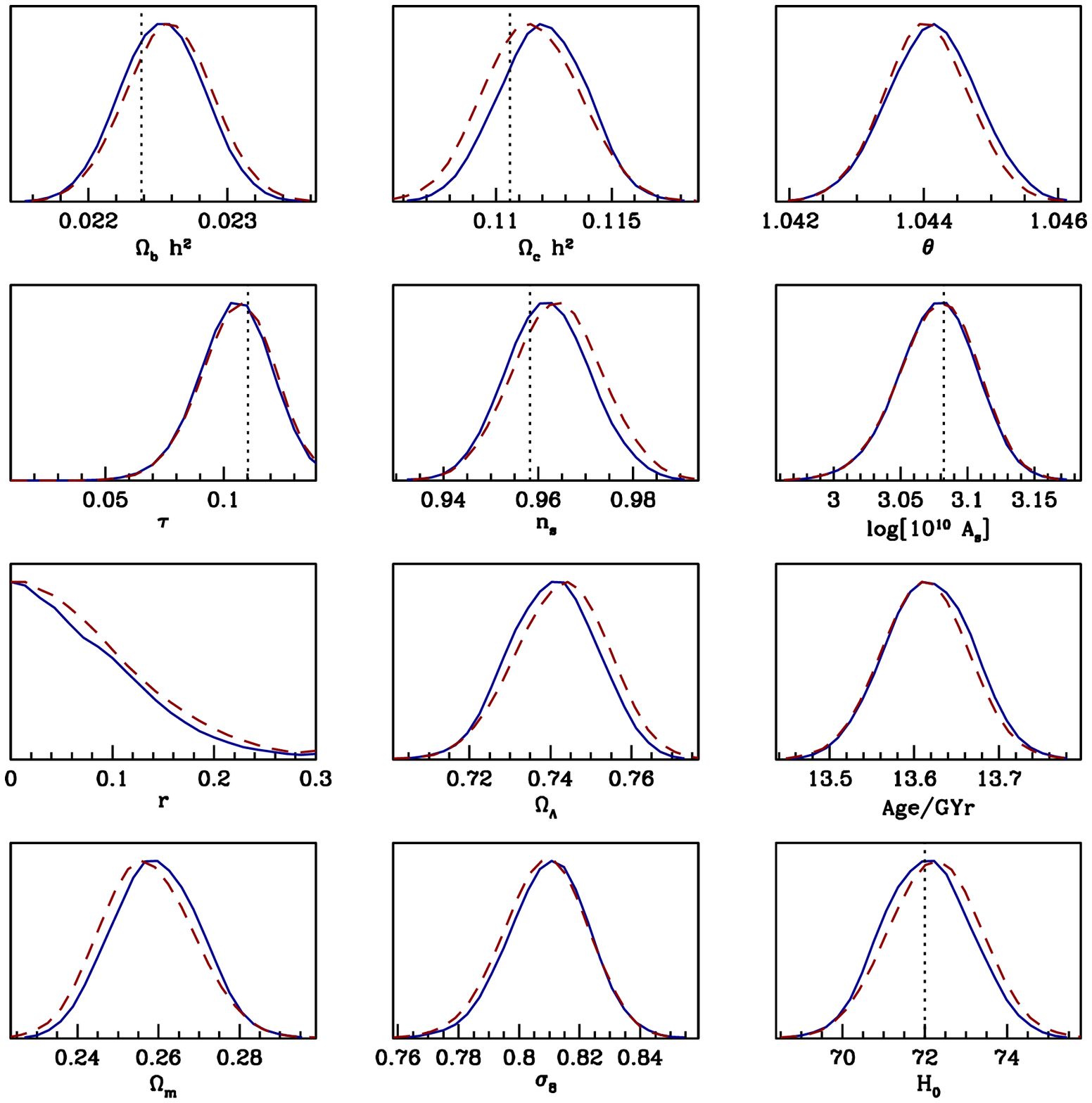}
}
\caption{Same as Figure~\ref{par-Offlogn-symm-asymm-obs}, but for the ensemble average of the Monte Carlo simulations.} 
\label{par-Offlogn-symm-asymm-avg}
\end{figure}

Figures~\ref{par-Offlogn-symm-asymm-obs} and \ref{par-Offlogn-symm-asymm-avg} show parameter distributions from the Offset Lognormal Bandpower likelihood in the symmetric and asymmetric beam cases, for the observed power spectrum and the ensemble average of the Monte Carlo simulations, respectively.  Parameters for the two cases are roughly consistent with each other.  They can be compared to the equivalent distributions from {\tt XFaster} likelihood in Figure~21 of \citet{XFaster-like2}.

Investigating the plot for the average mode, for the case {\tt XFaster} likelihood  we see deviations of the order of $\sigma/2$ for $\Omega_{c} h^{2}$, $\sigma_{8}$, $n_{s}$ and $H_{0}$. There is an obvious degeneracy between $\sigma_{8}$ and $n_{s}$.  For the observed case these deviations are noticeable mostly in $A_s$ and $\sigma_{8}$. Once again these parameters are degenerate.

The overall agreement  in the parameter constraints from both symmetric and asymmetric beams is quite impressive. This reflects the adequacy of our procedure when dealing with beam asymmetries, although there is still a slight bias for the asymmetric beam case. This bias is consistent with the small differences in the estimated parameters obtained with the Offset Lognormal Bandpower likelihood, in particular for parameters such as $n_{s}$, $\sigma_{8}$ and $log[10^{10} A_s]$. Although the power spectra look consistent, the parameter estimation shows departures of the order of $\sigma/2$ for some of the parameters.


 {\tt XFaster} assumes that the noise is white (uncorrelated), i.e., that the noise covariance matrix is diagonal.  Also, the {\tt XFaster} likelihood is estimated multipole by multipole, unlike the Offset Lognormal Bandpower likelihood. Hence a proper estimation of  the transfer function (filter function) requires a large number of Monte Carlo simulations to reduce the correlations between multipoles introduced by, e.g., sky cuts.  These simulations include both correlated noise and a sky cut. Hence for the small set of 100 Monte Carlos used here we should expect a larger deviation when {\tt XFaster} likelihood is employed.

\begin{figure}
\begin{center}
\hglue -5pt
\vbox{
\includegraphics[width=17.7cm]{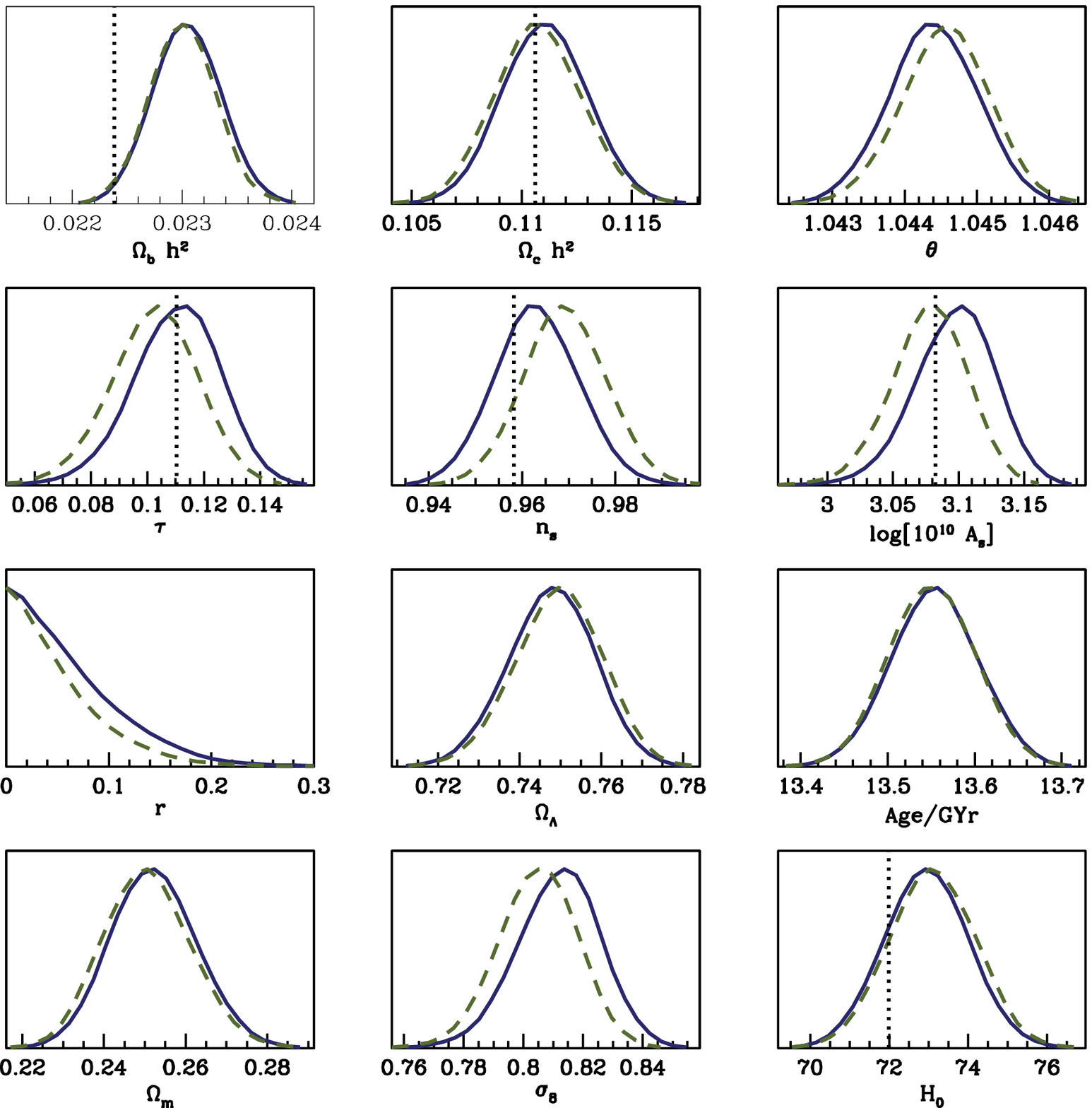}
}
\caption{Comparison of different window functions with the Offset Lognormal Bandpower likelihood.  Parameter distributions are shown for the observed map for top-hat (green dashed lines) and a Fisher (${\cal{F}}_{b b}$) (dark blue solid lines) window functions, for the asymmetric beam case.  Input model parameter values are marked by  vertical black lines.  Most parameters improve with  ${\cal{F}}_{b b}$-windows, while uncertainties are unaffected except in the case of $r$.   The 95\% upper limit on $r$ is higher by 15\% when using ${\cal{F}}_{b b}$ windows.}
\label{window-top-fbb}
\end{center}
\end{figure}

\begin{figure}
\hspace*{-1.2cm}
\includegraphics[width=19.75cm]{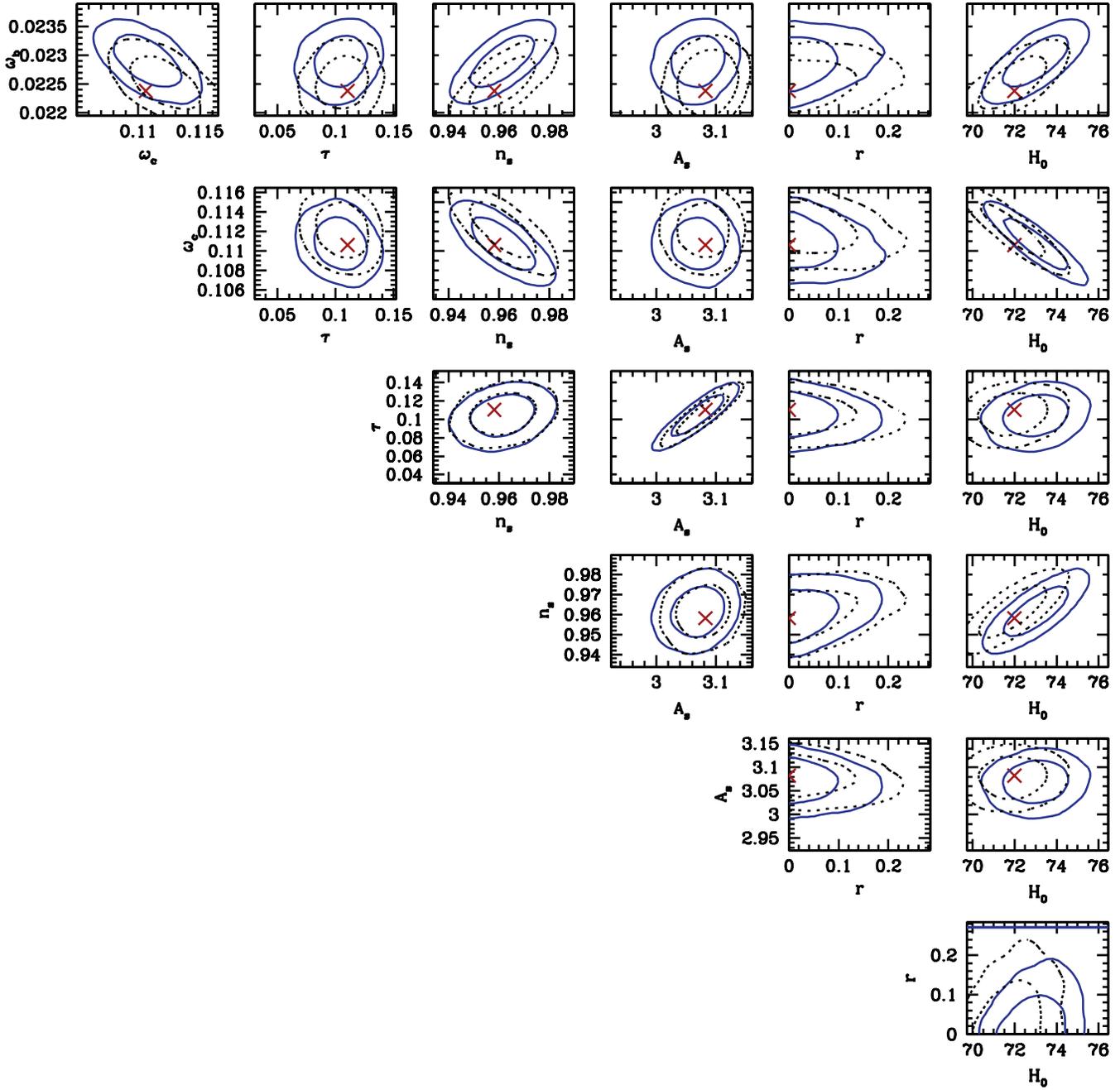}
\caption{Parameter constraints from the Offset Lognormal Bandpower likelihood with a top-hat window function, computed for the observed power spectrum (black dashed lines) and for the ensemble average of 100 Monte Carlo simulations (blue solid lines) }
\label{like2_box}
\end{figure}

\begin{figure}
\includegraphics[width=17.75cm]{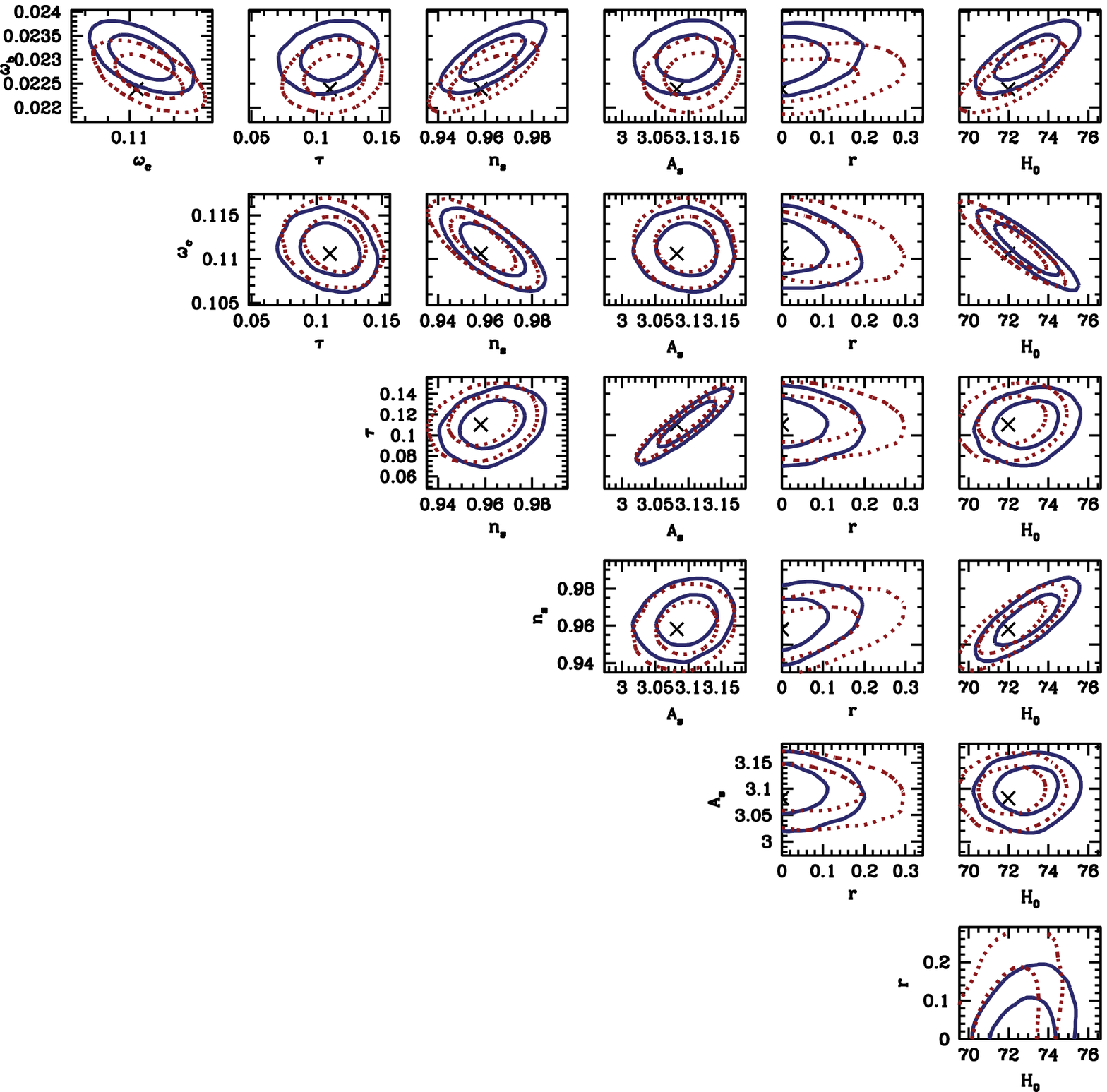}
\caption{Same as Figure~\ref{like2_box} for a Fisher (${\cal{F}}_{b b}$) window function, computed for the observed power spectrum (blue solid lines) and for the ensemble average of 100~Monte Carlo simulations (red dashed lines).}
\label{like2_fisher}
\end{figure}

Figures~\ref{window-top-fbb}, \ref{like2_box}, and \ref{like2_fisher} compare the effect of top-hat and Fisher (${\cal{F}}_{b b}$) window functions on parameters estimated with the Offset Lognormal Bandpower likelihood.  Fisher-weighted window functions account for the band power spectrum correlations.  It is clear that most parameters improve with ${\cal{F}}_{b b}$ windows, while uncertainties for all but $r$ are unaffected. The 95\% upper limit on $r$ is higher by 15\% for ${\cal{F}}_{b b}$ windows.

\citet{XFaster-like2} give a table of parameter constraints obtained with {\tt XFaster} likelihood for the symmetric beam case.  Table~\ref{par-wmap-planck-fi} is the same for the Offset Lognormal Bandpower likelihood and the {\tt XFaster} likelihood for the ensemble average power spectrum of the Monte Carlo simulations.
Our aim is solely to check how the parameter uncertainties for Planck from both likelihoods compare to those of WMAP \citep{Dunkley2009} and to our Fisher predictions.
The uncertainties are 2--3 times better than those for WMAP except for $\tau$  estimated with the Offset Lognormal Bandpower likelihood, for which the uncertainty on $\tau$ is 0.015 compared to $\sim$0.007 for {\tt XFaster}, 0.017 for WMAP, and 0.004 for our Fisher predictions (\citet {alpha}, Table~\ref{par-wmap-planck-fi}).

\begin{table*}
\caption{\label{par-wmap-planck-fi} Parameter estimates and uncertainties from the Offset Lognormal Bandpower and {\tt XFaster} likelihoods for Planck simulations at 70\,GHz, compared to WMAP \citep{Dunkley2009} and Fisher uncertainties (\citet{alpha}).  Estimates are for the ensemble average of 100 Monte Carlo simulations. Last column displays input parameter values $\pm$ Fisher uncertainties for reference.}
$$\hss\vbox{
   \newdimen\digitwidth 
   \setbox0=\hbox{\rm 0} 
   \digitwidth=\wd0 
   \catcode`*=\active 
   \def*{\kern\digitwidth}
   \newdimen\signwidth 
   \setbox0=\hbox{+} 
   \signwidth=\wd0 
   \catcode`!=\active 
   \def!{\kern\signwidth}
\halign{\hbox to 2cm{#\leaderfil}\tabskip=2em&
    \hfil#\hfil&
    \hfil#\hfil&
    \hfil#\hfil&
    \hfil#\hfil\tabskip=0pt\cr
\noalign{\doubleline}
\omit\hfil Parameter&WMAP&Offset Lognormal Bandpower&XFaster&Fisher\cr
\noalign{\vskip 3pt\hrule\vskip 5pt}
$\tau$&$0.087\pm0.017$&$0.099\pm0.015$&$0.1105^{+0.00643}_{-0.00771}$&$0.1103\pm0.004$ ($4$\%)\cr
\noalign{\vskip 5pt}
$n_s$ &$0.963\pm0.015$&$0.965\pm0.008$&$0.9621^{+0.01130}_{-0.01170}$&$0.9582\pm0.004$ ($0.4$\%)\cr
\noalign{\vskip 5pt}
$\omega_b$&$0.02273\pm0.015$&$0.0229\pm0.0003$&$0.0225\pm0.00042$&$0.02238\pm0.00018$ ($0.8$\%)\cr
\noalign{\vskip 5pt\hrule\vskip 3pt}}}
\hss$$\end{table*}


\section{Conclusions}

Parameters estimated with the {\tt XFaster} and Offset Lognormal Bandpower likelihoods agree well.  As the {\tt XFaster} likelihood is estimated for individual multipoles, a large number of Monte Carlo simulations is required for accurate estimates of low-$\ell$ correlations.  If only a small number of Monte Carlo simulations (such as the 100 used in this study), binning of the band power spectrum estimated with {\tt XFaster} used along with the Offset Lognormal Bandpower likelihood helps to partially correct these correlations.  For a large number of Monte Carlo simulations this is unnecessary. 
There {\tt XFaster} likelihood, however, has at least three advantages.  First, the Offset Lognormal likelihood does not properly take into account the temperature-polarization cross power.  This is likely to become evident with a larger number of simulations, or at lower noise levels, such as anticipated with the Planck HFI 143\,GHz channel.  We are investigating this further; results will be presented in \citet{GRhfiXFaster10}.  Second, the Offset Lognormal Bandpower likelihood requires calculation of a window function.  Third, the {\tt XFaster} likelihood can go straight from maps to parameters (via its raw pseudo-$C_{\ell}$), bypassing the band power spectrum estimation step. These advantages make {\tt XFaster} a adequate procedure to estimate cosmological parameters from Planck data in the high multipole regime.  As a bonus, {\tt XFaster} performs reasonably well for moderately low multipoles as well.  Although hybridization with a likelihood code able to handle fully the challenges of multipoles less than, say, 40, will be necessary for the best estimates parameters, {\tt XFaster} could be used alone where accuracy can be traded for speed.

\section{Acknowledgments}
The work reported in this paper was partially done within the CTP Working Group of
the {\sc Planck} Consortia. {\sc Planck} is a mission of the European Space Agency.
GR would like to thank useful discussions with Jeff Jewell.
This research used resources of the National Energy Research Scientific Computing Center, which is supported by the Office of Science of the U.S. Department of Energy under Contract No. DE-AC03-76SF00098.
This work has made use of the {\tt HEALPix} package \citep{healpix1}, and of the  {\sc Planck} satellite simulation package, {\tt LevelS,} \citep{levels}, which is assembled by the Max Planck Institute for Astrophysics {\sc Planck} Analysis Centre (MPAC).
The {\sc Planck} Project in the US is supported by the NASA Science Mission Directorate.
The research described in this paper was partially carried out at the Jet propulsion Laboratory, California Institute of Technology, under a contract with NASA.
Copyright 2009. All rights reserved.



\end{document}